\documentstyle[11pt,paspconf,psfig,twoside]{article}
\input{gcbook.pgi}
\begin{document}

\title{Radio Variability of Sgr A* at Centimeter Wavelengths}
\author{H. Falcke}
\affil{Max-Planck-Institut f\"ur Radioastronomie, Auf dem H\"ugel 69,
D-53121 Bonn, Germany}
\affil{Steward Observatory, The University of Arizona, Tucson, AZ 85721}

\begin{abstract}
Results of two years of continuous monitoring of flux density
variations at 8.3 and 2.3 GHz of the Galactic Center super-massive
black hole candidate Sgr A* are reported. The average RMS modulation
indices are 6\% and 2.5\% at 8.3 \& 2.3 GHz respectively. There is a
certain degree of correlation between both frequencies. The timescale
of variability at 8.3 \& 2.3 GHz is between 50 and 200 days. We
cannot confirm a $\lambda^2$ dependence of the timescale. At 2.3 GHz
a quasi-periodic behaviour with a period of 57 days was discovered
which is reminiscent to, though longer than, those found in some
compact extragalactic radio sources.
\end{abstract}


\section{Introduction}
Sgr~A* is believed to be the radio source associated with the
$2.6\cdot10^6 M_\odot$ (Haller et al.~1996; Ghez et al.~1998 \& 1999;
Eckart \& Genzel 1996; Genzel \& Eckart 1999; Zhao \& Goss~1999) dark mass
concentration in the center of the Galaxy. Since we know very little
about this source from other wavelengths, where it is extremely faint
(see Falcke 1996 for a review), a detailed study of its radio
properties is an important prerequisite for its interpretation. The
overall shape of the Sgr A* radio spectrum has been discussed in many
papers (e.g., Serabyn et al.~1997; Falcke et al.~1998) and the
variability has been investigated by Zhao et al.~(1989 \& 1992). The
spectral index ($S_\nu\propto\nu^\alpha$) of the source tends to be in
the range $\alpha\simeq0.2-0.3$ with an increasing value of $\alpha$ at
mm-wavelength and a possible cut-off at lower frequencies. At high
frequencies the spectrum cuts off in the infrared. A major problem
with the investigation of its radio variability is that Sgr A* is at
relatively low elevation for most interferometers, that it is embedded
in a large confusing structure, and that it becomes scatter-broadened
at low frequencies. The confusion especially is a major problem for
single-baseline interferometers with short baselines like the Green
Bank Interferometer (GBI) that is often used for variability
studies. For this reason the exact nature of the variability of Sgr~A*
has remained inconclusive. Flux density variations are clearly seen
between different epochs, but the timescale of the variability at
various frequencies is not well determined and it is not clear whether
some of the more extreme claims of variability are real or
instrumental artifacts. So far, Zhao et al.~(1989,1992) probably have
presented the largest database of Sgr A* flux-density
measurements. They found a number of outbursts at higher frequencies
and tentatively concluded that the small-amplitude variability at
longer wavelengths is caused by scattering effects in the ISM while
the variability at higher frequencies is intrinsic. In this paper new
results of a continuous monitoring program of Sgr A* at cm-wavelengths
performed with the GBI are presented and evaluated.

\section{GBI Observations and Data Reduction}
Sgr~A* has been part of the NASA/NRAO Green Bank Interferometer (GBI)
monitoring program for the past two years. The GBI is a two-element
interferometer (26m dishes) with a separation of 2400 meters,
operating simultaneously at X- and S-band (8.3 \& 2.3 GHz) with 35
MHz bandwidth. The resolution of the pencil beam is 3 and 11
arcseconds and 1 $\sigma$ noise levels are typically 30 and 6 mJy at X
and S-band respectively. The data are publically available but need
further processing, since the baseline gains depend on hourangle. In
addition observations of Sgr A* will also suffer from severe confusion
due to the small baseline and the extended structure of Sgr A West as
mentioned in the introduction.

The data were post-processed in the following way: an hourangle
dependent gain correction was fitted to 1622-297 which serves as a
calibrator to Sgr A*. Absolute gains were obtained using 3C286 as the
primary flux density calibrator. This gain corrections were then
applied to all sources and outliers were clipped when flux density
measurements deviated by more than 3 $\sigma$ from the median flux
density within a 20 day interval. For some calculations the data were
further averaged and gridded in three-day intervals. Only data after
July 1997 were considered due to initial calibration problems with the
GBI. All subsequent observations were made at almost the same hour
angle.

Sgr A* was also corrected for confusion. Comparison of the GBI data
with contemporaneous observations of Sgr A* at 5 and 8 GHz with the
VLA and VLBA (Bower et al.~1999a; Lo et al.~1998; Goss 1998, p.c.) were used
to calculate the difference between the GBI-single baseline flux
density and the total flux density of Sgr A*, where the 2.3 GHz total
flux density was obtained by extrapolation. Thus for an hourangle of
$\sim$ 0.88 hrs a flux of 70 and 177 mJy was added to the X and S-band
data respectively.

\begin{figure}[h]
\centerline{\psfig{figure=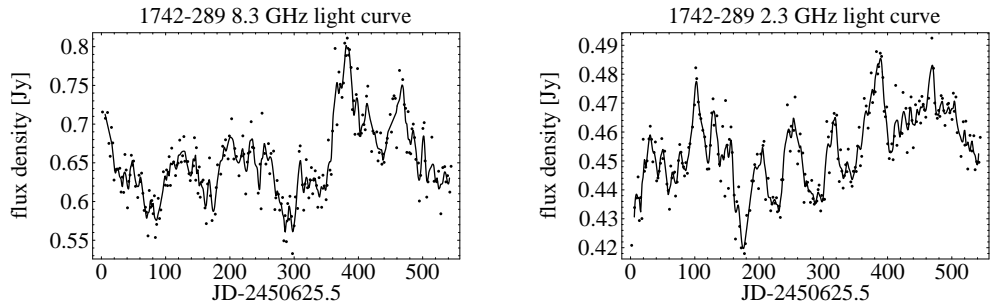,width=0.98\textwidth,bbllx=3.1cm,bblly=11.7cm,bburx=18.4cm,bbury=16.3cm}}
\caption[]{\label{lightcurve}Radio light curves of Sgr A* at 8.3 GHz (left panel) and
2.3 GHz (right panel) measured with the GBI (dots). The solid line is
the interpolated light curve for three-day averages.}
\end{figure}

The final light curves are shown in Figure~1. One can
see a peak-to-peak variability of 250 mJy and 60 mJy with an RMS of
6\% and 2.5\% at 8.3 \& 2.3 GHz, respectively (i.e.,~modulation
index). The median spectral index between the two frequencies for the
whole period is $\alpha=0.27$ ($S_\nu\propto\nu^\alpha$), varying
between 0.2 and 0.4.  There is a trend for the spectral index to
become larger when the flux density in both bands increases.

\section{Results}
To characterize the variability pattern better, Fig.~2
shows the structure function $D(\tau)$ of the two lightcurves, where
\begin{equation}
D(\tau)=\sqrt{\left<\left(S_\nu(t)-S_\nu(t\pm\tau)\right)^2\right>}\quad.
\end{equation}

A maximum in the structure function indicates a characteristic
timescale, a minimum indicates a characteristic period. A
characteristic period in radio-lightcurves usually does not persist
for a long time, and hence, similar to X-ray astronomy, is commonly
called a quasi-periodicity, even though the underlying physical
processes are probably very different from those seen in X-ray
binaries.

Interestingly, the structure functions at both frequencies look very
differently. While at both frequencies the characteristic time scale
is somewhere between 50 and 200 days, we find a clear signature of
quasi-periodic variability at 2.3 GHz, which is not obvious at 8.3
GHz. All the three maxima and the two minima in the structure function
are consistent with a period of 57 days.

\begin{figure}
\centerline{\psfig{figure=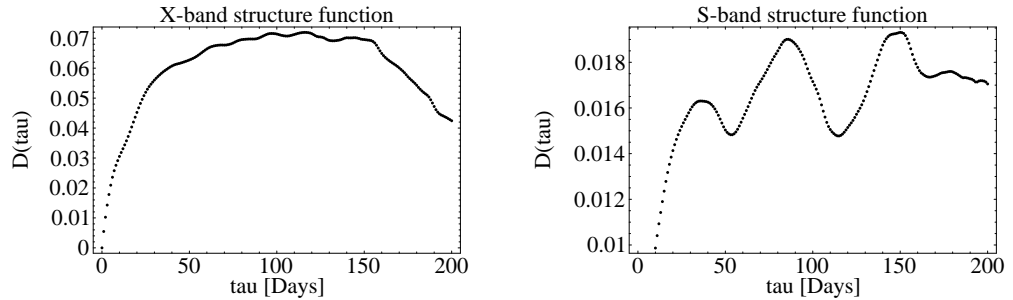,width=0.98\textwidth,bbllx=3.1cm,bblly=11.7cm,bburx=18.4cm,bbury=16.3cm}}
\caption[]{\label{structuref}Structure function of the radio light curves of Sgr A* at
8.3 GHz (left panel) and 2.3 GHz (right panel). Maxima indicate a
characteristic times scale, minima indicate a characteristic period}
\end{figure}

A cross correlation of the two light curves gives a strong peak near
zero time-lag which indicates a certain degree of correlation between
the emission at 8.5 GHz and 2.3 GHz (Fig.~3). A slight
offset of the peak by 2-3 days is visible (Fig.~3, right
panel). Usually such an offset would indicate that the 8.5 GHz light
curve precedes the one at 2.3 GHz. This would be qualitatively
expected by a model where outbursts travel outwards, from high to low
frequencies as for example in a jet model (Falcke et al.~1993),
however, the time lag one obtains is also close to the sampling rate
and it is not clear how significant this offset really is. Another
noteworthy feature of the cross correlation is that the 2.3 GHz
quasi-periodicity can still be seen. This could indicate that the
quasi-periodicity is also present at 8.3 GHz, but is swamped by
another, more erratic type of variability.

\begin{figure}
\centerline{\psfig{figure=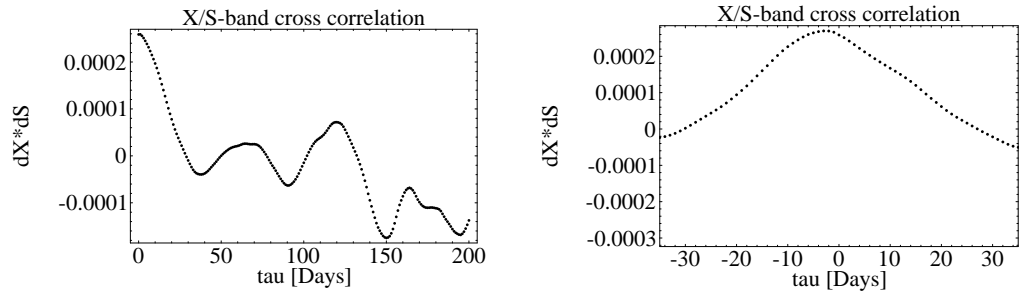,width=0.98\textwidth,bbllx=3.1cm,bblly=11.7cm,bburx=18.4cm,bbury=16.3cm}}
\caption[]{\label{cross}Cross correlation of 8.3 GHz and 2.3 GHz radio light curves
of Sgr A*. In the left panel positive and negative time lags have been
co-added. The right panel shows a blow-up of the cross correlation
(without co-adding positive and negative lags) around zero lag.}
\end{figure}

\section{Discussion and Summary}
To summarize the results one can say that there is clear evidence for
variability of a few percent at cm wavelengths in Sgr A*. The
variability does not seem to be consistent with a simple model of
refractive interstellar scintillation (RISS) as suggested by Zhao et
al.~(1989\&1992). The timescales at 2.3 GHz and 8.3 GHz both seem to
be comparable to the one found at 5 GHz by Zhao et al.~(1989\&1992)
and does not follow a $t\propto\lambda^2$ law. Moreover, the
modulation index apparently decreases towards lower frequencies.

The quasi-periodicity is reminiscent to those in some quasar
cores. For example the QSO 0917+624 is know to show episodes of
quasi-periodicity (Kraus et al.~1999). Unfortunately the frequency of
these quasi-periodicities in quasar cores and perhaps also Sgr A*, may
not be related to a well defined and constant (e.g., precession)
frequency like the QPOs in x-ray binaries, but could simply be due to
intermittent periodic phenomena in the accretion disk (e.g., waves) or
the jet (e.g., helical motion). In the case of Sgr A* all
characteristic timescales associated with a black hole or a
relativistic outflow at these frequencies are less than a day and
hence one might consider global accretion flow instabilities for such
a behaviour. On the other hand the possibility whether the
quasi-periodicity could be produced by interstellar scattering needs
to be explored as well.

\acknowledgments Helpful discussions with A. Kraus are gratefully
acknowledged. W.M. Goss provided VLA data for calibration
purposes. This work was supported by the Deutsche
Forschungsgemeinschaft, grants Fa 358/1-1\&2.  The Green Bank
Interferometer is a facility of the National Science foundation
operated by NRAO with support from the NASA High Energy Astrophysics
program.

\end{document}